\documentclass[american,aps,prb,amsmath,showpacs,twocolumn]{revtex4}
\usepackage{bm}
\usepackage{amsmath}
\usepackage{amssymb}
\usepackage{graphicx}

\renewcommand{\Re}{\mathop{\mathrm{Re}}}
\renewcommand{\Im}{\mathop{\mathrm{Im}}}

\usepackage{babel}
\begin{document}
\global\long\def\Pv{\mathop{\mathrm{P}}}

\title{New quantum-mechanical phenomenon in a model of electron-electron
interaction in graphene.}

\author{R.N. Lee}

\email{R.N.Lee@inp.nsk.su}

\author{A.I. Milstein}

\email{A.I.Milstein@inp.nsk.su}

\author{I.S. Terekhov}

\email{I.S.Terekhov@inp.nsk.su}

\affiliation{Budker Institute of Nuclear Physics of SB RAS, Novosibirsk, 630090
Russia}

\affiliation{Novosibirsk State University, Novosibirsk, 630090 Russia}

\begin{abstract}
A quantum mechanical model of two interacting electrons in graphene
is considered. We concentrate on the case of zero total momentum of
the pair. We show that the dynamics of the system is very unusual.
Both stationary and time-dependent problems are considered. It is
shown that the complete set of the wave functions with definite energy
includes the new functions, previously overlooked. The time evolution
of the wave packet, corresponding to the scattering problem setup,
leads to the appearance of the localized state at large time. The
asymptotics of this state is found analytically. We obtain the lower
bound of the life time of this state, which is connected with the
breakdown of the continuous model on the lattice scale. The estimate
of this bound gives one a hope to observe the localized states in
the experiment.
\end{abstract}
\maketitle

\section{Introduction}

Nowadays a great deal of effort has been devoted to the experimental
investigation of the transport properties of graphene, see recent
review \cite{Cooper2011}. One of the important results of these experiments
is the observation of high mobility of the charge carriers \cite{Novoselov2004}.
Many papers have been devoted to the investigation of the influence
of the electron-impurity interaction on the mobility of the charge
carriers, see reviews \cite{Kotov2011,DasSarma2011}. Study of the
electron-electron interaction in graphene is also important for understanding
of this effect, see Ref. \cite{Elias2011}. However, a theoretical
progress in this problem is rather limited \cite{Kotov2011,DasSarma2011}.

It is well established now that the low-energy single electron dynamics
in graphene is described by a massless two-component Dirac equation
\cite{Wallace1947,McClure1956,Gonzalez1993,Gonzalez1994,Novoselov2004}
\begin{gather*}
i\hbar\partial_{t}\psi\left(t,\mathbf{r}\right)=\hat{h}\psi\left(t,\mathbf{r}\right),
\end{gather*}
where the hamiltonian $\hat{h}$ has the form
\[
\hat{h}=v_{F}\bm{\sigma}\cdot\hat{\mathbf{p}},
\]
$v_{F}$ is the Fermi velocity, $\hat{\mathbf{p}}=-i\hbar\bm{\nabla}$,
and $\bm{\sigma}=\left(\sigma_{x},\sigma_{y}\right)$ are the Pauli
matrices acting on the pseudospin variables. Below we set $\hbar=v_{F}=1$.
Evidently, the pair of non-interacting electrons can be described
by the equation
\begin{eqnarray}
i\partial_{t}\psi\left(\mathbf{r}_{1},\mathbf{r}_{2},t\right) & = & \hat{H}_{0}\psi\left(\mathbf{r}_{1},\mathbf{r}_{2},t\right)\,,\label{eq:Free}\\
\hat{H}_{0}=\hat{h}_{1}+\hat{h}_{2} & = & \bm{\sigma}_{1}\cdot\hat{\mathbf{p}}_{1}+\bm{\sigma}_{2}\cdot\hat{\mathbf{p}}_{2}\,,
\end{eqnarray}
where $\psi\left(\mathbf{r}_{1},\mathbf{r}_{2},t\right)$ is the wave
function of the system, depending on the coordinates and pseudospin
variables of both electrons. The generalization of Eq. \eqref{eq:Free}
to the case of interacting electrons is a highly nontrivial problem.
The origin of the difficulties is the necessity to take into account
the interaction with the electrons below Fermi surface. This interaction
results in the existence of the electron-hole excitations in the intermediate
states. The account of the corresponding effects in quantum electrodynamics
(QED) leads to the Dyson-Schwinger equation (which, for the bound
states, reduces to the Bethe-Salpeter equation), see, e.g., Ref.\cite{Berestetski1982}.
However, in the nonrelativistic QED systems, the effect of virtual
electron-positron pair in the intermediate states is small. For massless
electrons in graphene, the nonrelativistic approximation is not applicable
and the effect of virtual electron-hole excitation may be crucially
important for the problem of electron-electron interaction. The approach
based on the Bethe-Salpeter equation was used in Ref. \cite{Gamayun2009}
in the investigation of electron-hole interaction in graphene.

Though the influence of the electron-hole excitations can be very
important, nevertheless, as a first step in the investigation of the
electron-electron interaction, it makes sense to ignore this effect
and to model the electron-electron interaction by replacing $\hat{H}_{0}\to\hat{H}_{V}$
in Eq. \eqref{eq:Free}, where
\begin{equation}
\hat{H}_{V}=\hat{H}_{0}+V\left(r\right)=\bm{\sigma}_{1}\cdot\hat{\mathbf{p}}_{1}+\bm{\sigma}_{2}\cdot\hat{\mathbf{p}}_{2}+V\left(r\right)\label{eq:H}
\end{equation}
and $V\left(r\right)=V(\left|\bm{r}_{1}-\bm{r}_{2}\right|)$ is the
electron-electron interaction potential. Recently, this model has
been considered in Ref.\cite{Sabio2010}, where the eigenfunctions
of $\hat{H}_{V}$ have been analysed. The solutions found in Ref.\cite{Sabio2010}
appeared to have unusual properties. In order to understand the origin
of these properties, we revisit in the present paper the solution
of the stationary equation $\hat{H}_{V}\psi=E\psi$. We also consider
the time-dependent problem and demonstrate that the unusual properties
of the eigenfunctions of the hamiltonian are reflected in the very
specific properties of the time evolution of the wave packets.

We restrict our consideration to the specific case of zero total momentum
of the pair and search for the solutions being the eigenstates of
the operator
\[
\hat{J}^{z}=\frac{1}{2}\left(\sigma_{1}^{z}+\sigma_{2}^{z}\right)-i\partial_{\varphi}\,,
\]
where $\varphi$ is the azimuth angle of the vector $\mathbf{r}=\mathbf{r}_{1}-\mathbf{r}_{2}$.
We assume that the potential $V\left(r\right)$ is a smooth positive
monotonically decreasing vanishing function. To include the important
case of the Coulomb potential into the consideration, we allow for
the $r^{-1}$ growth of the potential at $r\to0$. The solution of
the stationary equation shows that the wave functions with $J^{z}=0$
are smooth functions for any energy $E$. This is also valid for the
wave functions with $J_{z}\neq0$ and the energy above the maximum
of the potential $V_{\mathrm{max}}=V\left(0\right)$ or below zero.
For $J_{z}\neq0$ and $0<E<V_{\mathrm{max}}$, the solution of the
stationary equation necessarily has singularity at the point $r_{\star}\left(E\right)$
determined by the condition
\begin{equation}
E=V\left(r_{\star}\right),\label{eq:r_star}
\end{equation}
which is in agreement with Ref. \cite{Sabio2010}. Such a behaviour
contradicts a common wisdom which tells one that the wave function
should be a smooth function in the region where the potential is also
smooth. We show that the existence of the singularity in the wave
function is related to the degeneracy of the derivative matrix in
the hamiltonian. We find an important new feature of the energy spectrum:
the additional degeneracy of the states with fixed $J^{z}\neq0$ and
energy in the interval $\left(0,V_{\mathrm{max}}\right)$.

For the time-dependent problem we choose the initial conditions corresponding
to the wide spherical wave packet with fixed $J^{z}$ and the average
energy $E_{0}$ (average value of the hamiltonian) moving toward the
origin from the large distance $r_{0}\gg\Delta$, where $\Delta$
is the width of the packet (the energy dispersion in the packet $\sim\Delta^{-1}$).
The direct numerical calculation reveals a remarkable picture. At
rather large time $t\gtrsim r_{0}$ one observes not only a reflected
wave packet moving toward the large $r$, but also a narrow peak in
the vicinity of $r_{\star}\left(E_{0}\right)$ with the width $\propto1/\Delta$.
The total norm of the wave function is conserved as it should be for
a hermitian hamiltonian. In order to check consistency of the results
obtained, we demonstrate that the solution of the time-dependent equation
based on the decomposition of the initial wave packet over the stationary
wave functions reproduces the direct numerical solution of this equation.
Using this decomposition, we find the large-time asymptotics of the
emerged peak analytically.

\section{General properties of the model}

Obviously, the hamiltonian $\hat{H}$ in Eq. \eqref{eq:H} commutes
with the total momentum $\mathbf{P}=\mathbf{p}_{1}+\mathbf{p}_{2}$,
and we can search the wave function in the form $\psi\left(\mathbf{r}_{1},\mathbf{r}_{2},t\right)=\exp\left(i\mathbf{P}_{0}\cdot\mathbf{R}\right)\psi\left(t,\mathbf{r}\right),$
where $\mathbf{R}=\left(\mathbf{r}_{1}+\mathbf{r}_{2}\right)/2$ is
the center-of-energy coordinate and $\mathbf{r}=\mathbf{r}_{1}-\mathbf{r}_{2}$
is the relative position vector. Note that the wave function $\psi\left(t,\mathbf{r}\right)$
depends nontrivially on the system total momentum $\mathbf{P}_{0}$,
see Ref. \cite{Sabio2010}. Below we consider a specific case $\mathbf{P}_{0}=0$.
Then the wave equation has the form
\begin{align}
 & i\partial_{t}\psi\left(t,\mathbf{r}\right)=\hat{H}\psi\left(t,\mathbf{r}\right),\label{eq:H[P=00003DP0]}\\
 & \hat{H}=\left(\boldsymbol{\sigma}_{1}-\boldsymbol{\sigma}_{2}\right)\cdot\hat{\mathbf{p}}+V\left(r\right)\,,
\end{align}
where $\hat{\mathbf{p}}=-i\bm{\nabla}$. The hamiltonian $\hat{H}$
commutes with the operator
\begin{equation}
\hat{J}^{z}=\hat{S}^{z}+\hat{L}^{z}=\frac{1}{2}\left(\sigma_{1}^{z}+\sigma_{2}^{z}\right)-i\partial_{\varphi}
\end{equation}
($\varphi$ is the azimuthal angle of the vector $\mathbf{r}$) and
the operator
\begin{equation}
\hat{O}=\hat{\mathbf{S}}^{2}-2\left(\hat{S}^{z}\right)^{2},\label{eq:operator O}
\end{equation}
 where $\hat{\mathbf{S}}^{2}=\frac{1}{4}\sum_{i=1}^{3}\left(\sigma_{1}^{i}+\sigma_{2}^{i}\right)^{2}$.
The operators $\hat{J}^{z}$ and $\hat{O}$ also commute with each
other. Therefore, we can search for the solution of Eq. \eqref{eq:H[P=00003DP0]}
to be the eigenfunction of $\hat{J}^{z}$ and $\hat{O}$:
\begin{align}
\psi_{0}\left(t,\mathbf{r}\right) & =e^{iM\varphi}\Bigl(a_{00}\left(t,r\right)\left|0,0\right\rangle +e^{-i\varphi}a_{11}\left(t,r\right)\left|1,1\right\rangle \nonumber \\
 & +e^{i\varphi}a_{1-1}\left(t,r\right)\left|1,-1\right\rangle \Bigr),\label{eq:psi_1}\\
\psi_{2}\left(t,\mathbf{r}\right) & =e^{iM\varphi}g\left(t,r\right)\left|1,0\right\rangle \label{eq:psi_2}
\end{align}
so that $\hat{J}^{z}\psi_{k}=M\psi_{k}$ and $\hat{O}\psi_{k}=k\psi_{k}$.
Here $\left|s,s_{z}\right\rangle $ is the eigenfunction of the operators
$\hat{\mathbf{S}}^{2}$ and $\hat{S}^{z}$. It is convenient to pass
from the functions $a_{ij}$ to the functions
\begin{equation}
f=i\frac{a_{11}+a_{1-1}}{\sqrt{2}}\,,\quad h=i\frac{a_{11}-a_{1-1}}{\sqrt{2}}\,,\quad d=a_{00}.\label{eq:fhdg definition}
\end{equation}
Using Eqs. \eqref{eq:H[P=00003DP0]}, \eqref{eq:psi_1}, \eqref{eq:psi_2},
and \eqref{eq:fhdg definition}, we obtain
\begin{align}
i\partial_{t}g & =V\left(r\right)g,\label{eq:g equation}\\
i\partial_{t}f & =V\left(r\right)f-\frac{2M}{r}d,\label{eq:f equation}\\
i\partial_{t}h & =V\left(r\right)h-2\partial_{r}d,\label{eq:h equation}\\
i\partial_{t}d & =V\left(r\right)d-\frac{2M}{r}f+2\left(\partial_{r}+\frac{1}{r}\right)h,\label{eq:d equation}
\end{align}
The last three equations can be represented in the matrix form

\begin{align}
i\partial_{t}F & =\hat{H}_{r}F,\label{eq:F equation}
\end{align}
where
\begin{equation}
F=\begin{pmatrix}f\\
h\\
d
\end{pmatrix}\,,\quad\hat{H}_{r}=\begin{pmatrix}V\left(r\right) & 0 & -\frac{2M}{r}\\
0 & V\left(r\right) & -2\partial_{r}\\
-\frac{2M}{r} & 2\left(\partial_{r}+\frac{1}{r}\right) & V\left(r\right)
\end{pmatrix}\,.\label{eq:Hr}
\end{equation}
It is easy to see that $\hat{H}_{r}$ is a hermitian operator, i.e.
\[
\intop_{0}^{\infty}drrF_{1}^{\dagger}\hat{H}_{r}F_{2}=\intop_{0}^{\infty}drr\left(\hat{H}_{r}F_{1}\right)^{\dagger}F_{2}
\]
 for continuous functions $F_{1,2}\left(r\right)$, decreasing sufficiently
fast when $r\to\infty$ and finite at $r=0$.

The general solution of Eq. \eqref{eq:g equation} is
\begin{equation}
g\left(t,r\right)=g\left(0,r\right)e^{-iV\left(r\right)t},\label{eq:g(t,r)}
\end{equation}
 whereas the general solution of Eq. \eqref{eq:F equation} can not
be found analytically.

\paragraph*{Conserved current and density.}

The conserved current and density for Eq. \eqref{eq:H[P=00003DP0]}
have the form

\begin{equation}
\mathbf{j}=\psi^{\dagger}\left(\boldsymbol{\sigma}_{1}-\boldsymbol{\sigma}_{2}\right)\psi\,,\quad\rho=\psi^{\dagger}\psi.\label{eq:current1}
\end{equation}
For two solutions $\psi_{0}$ and $\psi_{2}$, Eq. \eqref{eq:psi_2},
the current and density are expressed as
\begin{align}
\psi_{0}: & \quad j_{r}=4\Im\left(dh^{*}\right),\quad j_{\varphi}=-4\Re\left(df^{*}\right),\nonumber \\
 & \quad\rho=\left|f\right|^{2}+\left|h\right|^{2}+\left|d\right|^{2},\label{eq:Psi0Current}\\
\psi_{2}: & \quad j_{r}=0,\quad j_{\varphi}=0,\quad\rho=\left|g\right|^{2},\label{eq:Psi2Current}
\end{align}
where $j_{r}$ and $j_{\varphi}$ are the radial and angular components
of the current, respectively.

\section{Stationary problem}

Let us first consider the stationary equation for the function $g$:
\begin{equation}
Eg=V\left(r\right)g\,.\label{eq:stationary_g}
\end{equation}
This simple consideration helps one to understand better the properties
of the solutions of the stationary equation for the function $F$.
For $E>V_{\mathrm{max}}$, the equation \eqref{eq:stationary_g} has
no solutions, while for $0<E<V_{\mathrm{max}}$ its formal solution
is $g_{a}\left(r\right)=\delta\left(r-a\right)$, where $a$ is determined
by the equation $E=V\left(a\right)$. The functions $g_{a}\left(r\right)$
for different values of $a$ are mutually orthogonal and normalized
by the condition
\begin{equation}
\intop_{0}^{\infty}dr\, r\, g_{a}\left(r\right)g_{\tilde{a}}\left(r\right)=a\,\delta\left(a-\tilde{a}\right)\,.
\end{equation}
Note that the density $\rho\left(r\right)=\left|g_{a}\left(r\right)\right|^{2}$,
Eq. \eqref{eq:Psi2Current}, corresponding to this solution, is not
well-defined. Nevertheless, the functions $g_{a}\left(r\right)$ form
a complete set and can be used to solve the time-dependent problem.
Indeed,
\[
g\left(t,r\right)=\intop_{0}^{\infty}da\, e^{-iV\left(a\right)t}g\left(0,a\right)g_{a}\left(r\right)=g\left(0,r\right)e^{-iV\left(r\right)t}
\]
in agreement with Eq. \eqref{eq:g(t,r)}.

Let us now pass to the consideration of the stationary equation
\begin{equation}
EF=\hat{H}_{r}F\,.\label{eq:eF=00003DHF}
\end{equation}
The hamiltonian $\hat{H}_{r}$ is a first-order differential operator,
see Eq. \eqref{eq:Hr}. It is known from the theory of ordinary differential
equations that the solution $y$ of the system $\partial_{r}y\left(r\right)=A\left(r\right)y\left(r\right)$
can have singularities only in the points where the elements of the
matrix $A\left(r\right)$ are singular. We can not, however, represent
Eq. \eqref{eq:eF=00003DHF} in this form since the matrix $\begin{pmatrix}0 & 0 & 0\\
0 & 0 & -2\\
0 & 2 & 0
\end{pmatrix}$ in front of the derivative $\partial_{r}$ in $\hat{H}_{r}$ is degenerate.
We show below that this degeneracy leads, for $M\neq0$, to the appearance
of the singularity of the solution $F$ in the point $r=r_{\star}$.

\paragraph*{Second-order equation for $d$.}

The explicit form of Eq. \eqref{eq:eF=00003DHF} reads
\begin{align}
\left(E-V\right)f & =-\frac{2M}{r}d\,,\label{eq:f_stationary_equation}\\
\left(E-V\right)h & =-2\partial_{r}d\,,\label{eq:h_stationary_equation}\\
\left(E-V\right)d & =2\partial_{r}h+\frac{2}{r}h-\frac{2M}{r}f\,.\label{eq:d_stationary_equation}
\end{align}
Using the first two equations in order to eliminate the functions
$f$ and $h$ from the last equation, we obtain
\begin{gather}
d^{\prime\prime}+p\left(r\right)d^{\prime}+q\left(r\right)d=0,\label{eq:d'' equation}\\
p\left(r\right)=\frac{V^{\prime}}{E-V}+\frac{1}{r}\,,\label{eq:p(r)}\\
q\left(r\right)=\frac{1}{4}\left(E-V\right)^{2}-\frac{M}{r^{2}}^{2}\,,\label{eq:q(r)}
\end{gather}
where a prime denotes the derivative with respect to $r$.

\paragraph*{Boundary condition at $r=0$.}

Let us determine the boundary condition at $r=0$. If $V\left(0\right)<\infty$,
the general solution of Eq. \eqref{eq:d'' equation} behaves near
$r=0$ as
\begin{equation}
d\approx a_{1}r^{\left|M\right|}+a_{2}\times\begin{cases}
r^{-\left|M\right|}, & M\neq0\\
\ln r, & M=0
\end{cases}\,,
\end{equation}
while for the case of Coulomb singularity, when $V\left(r\right)\stackrel{r\to0}{\to}\alpha/r$,
the asymptotics of the general solution has the form
\begin{equation}
d\approx a_{1}r^{\nu-1/2}+a_{2}r^{-\nu-1/2}\,,\label{eq:d r=00003D0 asymptotics Coulomb}
\end{equation}
where $\nu=\frac{1}{2}\sqrt{4M^{2}+1-\alpha^{2}}$. Here $a_{1}$
and $a_{2}$ are some constants. We choose the boundary condition
at $r=0$ as
\begin{equation}
a_{2}=0\,.\label{eq:boundary_condition_at_0}
\end{equation}
This condition provides that $\int_{0}^{\delta}dr\, r\left|F\right|^{2}<\infty$
for sufficiently small $\delta>0$.

\paragraph*{Analytical properties.}

To understand the properties of the solution $d$ of Eq. \eqref{eq:d'' equation},
we consider the analytical properties of the coefficients $p\left(r\right)$
and $q\left(r\right)$, Eqs. \eqref{eq:p(r)}, \eqref{eq:q(r)}, on
the interval $\left[0,\infty\right)$. In the origin, the coefficients
behave as $p\left(r\right)\sim r^{-1}$, $q\left(r\right)\sim r^{-2}$,
so that the point $r=0$ is a regular singular point of the differential
equation \eqref{eq:d'' equation}. The coefficient $q\left(r\right)$
tends to a constant when $r\to\infty$, therefore the point $r=\infty$
is an irregular singular point of the equation. The above properties
of Eq. \eqref{eq:d'' equation} (singularities of $p\left(r\right)$
and $q\left(r\right)$ at $r=0,\infty$ and boundary condition for
$d\left(r\right)$ at $r=0$) are analogous to those of radial Schr\"odinger
equation. The new property of Eq. \eqref{eq:d'' equation} is the
singularity of the coefficient $p\left(r\right)$ at $r=r_{\star}\left(E\right)$,
see Eq. \eqref{eq:r_star}, when $E\in\left(0,V_{\mathrm{max}}\right)$.
In principle, the singularities of the coefficients of the equation
do not necessarily lead to the singularity of the solution (and its
derivatives). One can check that the general solution for $M=0$ is,
indeed, a smooth function at $r=r_{\star}$. Therefore, we concentrate
on the case $M\neq0$. The general solution of Eq. \eqref{eq:d'' equation}
in this case is not smooth at $r=r_{\star}$, which can be readily
seen from the asymptotics of the solution in the vicinity of $r_{\star}$:
\begin{align}
d\left(r\right) & \approx b_{1}\left(1+\frac{M^{2}\left(1-r/r_{\star}\right)^{2}}{2}\ln\left|1-r/r_{\star}\right|\right)\nonumber \\
 & +b_{2}\left(1-r/r_{\star}\right)^{2}\,.
\end{align}
Therefore, we search for the solution separately in two regions
\begin{equation}
d\left(r\right)=\begin{cases}
\tilde{b}_{1}d_{\mathrm{irr}}\left(r\right)+\tilde{b}_{2}d_{\mathrm{reg}}\left(r\right), & 0<r<r_{\star}\\
b_{1}d_{\mathrm{irr}}\left(r\right)+b_{2}d_{\mathrm{reg}}\left(r\right) & r_{\star}<r<\infty
\end{cases}\,,
\end{equation}
where $b_{1,2}$ and $\tilde{b}_{1,2}$ are some constants, and the
functions $d_{\mathrm{irr}}$ and $d_{\mathrm{reg}}$ have the asymptotics
\begin{eqnarray}
d_{\mathrm{irr}}\left(r\right) & \approx & 1+\frac{M^{2}\left(r_{\star}-r\right)^{2}}{2r_{\star}^{2}}\ln\left|1-r/r_{\star}\right|\,,\\
d_{\mathrm{reg}}\left(r\right) & \approx & \frac{\left(r_{\star}-r\right)^{2}}{r_{\star}^{2}}\,
\end{eqnarray}
at $r\to r_{\star}$.

\paragraph*{Matching conditions at $r=r_{\star}$.}

In order to determine the general form of the solution of Eq. \eqref{eq:d'' equation},
we need to apply matching conditions at $r=r_{\star}$. Conventional
analysis of Eq. \eqref{eq:d'' equation} in the vicinity of $r_{\star}$
leads to the requirement of the continuity of $d$ and $d^{\prime}$
at $r=r_{\star}$. Using these conditions we end up with
\begin{equation}
\tilde{b}_{1}=b_{1}\,.\label{eq:b1=00003Db1}
\end{equation}
Note that \textbf{$b_{2}$} remains a free parameter, which means
that the function $d\left(r\right)$ in the region $r>r_{\star}$
is not entirely determined by that in the region $r<r_{\star}$.

The boundary condition at the origin, Eq. \eqref{eq:boundary_condition_at_0},
fixes the ratio
\begin{equation}
\tilde{b}_{2}/\tilde{b}_{1}=\beta\,,\label{eq:b2/b1}
\end{equation}
where $\beta$ is the constant which depends on the energy and the
form of the potential. Therefore, we have two conditions, \eqref{eq:b1=00003Db1}
and \eqref{eq:b2/b1}, for four constants $b_{1,2}$ and $\tilde{b}_{1,2}$.
It means that for any energy in the interval $E\in\left(0,V_{\mathrm{max}}\right)$
there are two linearly independent solutions which we choose as
\begin{eqnarray}
d_{1}\left(r\right) & = & b_{1}\left[d_{\mathrm{irr}}\left(r\right)+\beta d_{\mathrm{reg}}\left(r\right)\right],\label{eq:d1_solution}\\
d_{2}\left(r\right) & = & b_{2}\theta\left(r-r_{\star}\right)d_{\mathrm{reg}}\left(r\right)\,,\label{eq:d2_solution}
\end{eqnarray}
where $\theta\left(x\right)$ is the Heaviside step function.

\paragraph*{Solutions of the system \eqref{eq:f_stationary_equation}--\eqref{eq:d_stationary_equation}.}

Let us now return to the initial system, Eqs. \eqref{eq:f_stationary_equation}--\eqref{eq:d_stationary_equation}.
Note that Eqs. \eqref{eq:f_stationary_equation} and \eqref{eq:h_stationary_equation}
determine functions $f$ and $h$ up to the generalized function localized
at $r=r_{\star}$. Substituting \eqref{eq:d1_solution} and \eqref{eq:d2_solution}
in Eqs. \eqref{eq:f_stationary_equation}--\eqref{eq:d_stationary_equation},
we obtain two solutions of the equation \eqref{eq:eF=00003DHF}:\begin{widetext}
\begin{eqnarray}
F_{1}\left(E,r\right) & = & \left(\begin{array}{c}
\frac{d_{1}\left(E,r\right)}{r}\Pv\frac{2M}{V\left(r\right)-E}\\
\frac{2\partial_{r}d_{1}\left(E,r\right)}{V\left(r\right)-E}\\
d_{1}\left(E,r\right)
\end{array}\right)\,,\label{eq:F1}\\
F_{2}\left(E,r\right) & = & \left(\begin{array}{c}
\frac{2Md_{2}\left(E,r\right)}{r\left(V\left(r\right)-E\right)}-\frac{2r\partial_{r}^{2}d_{2}\left(E,r_{\star}+0\right)}{M}\delta\left(V\left(r\right)-E\right)\\
\frac{2\partial_{r}d_{2}\left(E,r\right)}{V\left(r\right)-E}\\
d_{2}\left(E,r\right)
\end{array}\right)\,.\label{eq:F2}
\end{eqnarray}
\end{widetext}Here $\Pv\frac{1}{x}$ stands for the principal value
defined as
\[
\Pv\frac{1}{V\left(r\right)-E}=\frac{1}{2}\left(\frac{1}{V\left(r\right)-E+i0}+\frac{1}{V\left(r\right)-E-i0}\right)\,.
\]
In order to check that $F_{1}$ and $F_{2}$ are the solutions of
Eq. \eqref{eq:eF=00003DHF} in the vicinity of $r_{\star}$, one can
integrate the equations \eqref{eq:f_stationary_equation}--\eqref{eq:d_stationary_equation}
over $r$ from $r_{\star}-\delta_{1}$ to $r_{\star}+\delta_{2}$,
and consider the limit $\delta_{1,2}\to+0$.

Similar to the solutions $g_{a}\left(r\right)$ of the equation \eqref{eq:stationary_g},
the functions $F_{1,2}$ contain generalized functions. The density
$\rho=F^{\dagger}F$, corresponding to the solution $F_{1}$ is not
integrable at the point $r=r_{\star}$, while that, corresponding
to $F_{2}$, is ill-defined. Therefore, for energies in the interval
$\left(0,V_{\mathrm{max}}\right)$ there is no solution $F\left(E,r\right)$
of the stationary equation with the density being an integrable function
in the vicinity of $r_{\star}\left(E\right)$. This statement is in
clear contradiction with the statement of Ref. \cite{Sabio2010},
where it was claimed that nonanalyticites at $r=r_{\star}$ give a
finite contribution to the probability.

It may seem that the same consideration of the case $M=0$ will also
lead to the two-fold degeneracy of the spectrum for $0<E<V_{\mathrm{max}}$.
However, it turns out that the substitution of $d_{2}$ from Eq. \eqref{eq:d2_solution}
to the original system, Eqs. \eqref{eq:f_stationary_equation}--\eqref{eq:d_stationary_equation},
leads, for $M=0$, to appearance of the $\delta$-function term violating
Eq. \eqref{eq:d_stationary_equation}.

For completeness, let us also discuss the properties of the solution
of Eq. \eqref{eq:eF=00003DHF} for the energies above the maximum
of the potential or below zero. In this case there is no singularity
in the coefficient $p\left(r\right)$ on the interval $\left(0,\infty\right)$.
The energy spectrum is not degenerate for $E<0$ and $E>V_{\mathrm{max}}$
since the solution is defined uniquely (up to the normalization) by
the boundary condition at the origin. This solution has the form \eqref{eq:F1}
where one can omit the $\Pv$ symbol. In what follows we assume that
$F_{1}\left(E,r\right)$ for $E<0$ and $E>V_{\mathrm{max}}$ is normalized
as
\begin{equation}
\int_{0}^{\infty}dr\, rF_{1}^{\dagger}\left(E,r\right)F_{1}\left(E^{\prime},r\right)=2\pi\delta\left(E-E^{\prime}\right)\,,
\end{equation}
 so that the large-$r$ asymptotics of $F_{1}\left(E,r\right)$ has
the form
\begin{equation}
F_{1}\left(E,r\right)\stackrel{r\to\infty}{\longrightarrow}\frac{1}{\sqrt{r}}\left(\begin{array}{c}
0\\
\sin\frac{Er+\varphi}{2}\\
\cos\frac{Er+\varphi}{2}
\end{array}\right)\,,\label{eq:F1 asymptotics}
\end{equation}
where $\varphi$ is some function of the energy.

\paragraph*{Alternative derivation.}

We present now an alternative derivation of the solutions \eqref{eq:F1}
and \eqref{eq:F2} for $M\neq0$, which allows one to understand better
the appearance of the second solution $F_{2}$. For this purpose we
interpret Eq. \eqref{eq:eF=00003DHF} for real $E$ as a limit of
the equation
\begin{equation}
\left(E+i\epsilon\right)F=\hat{H}_{r}F\label{eq:(E+ie)F=00003DHF}
\end{equation}
 at $\epsilon\to\pm0$. The limit depends on the sign of $\epsilon$
(see below) and we denote the corresponding solutions by the lower
index $+$ or $-$, respectively. The equation for the function $d\left(r\right)$
has the form \eqref{eq:d'' equation} with the replacement $E\to E+i\epsilon$.
For $\epsilon\neq0$, the coefficient $p\left(r\right)$ of this equation
has no singularities on the interval $\left(0,\infty\right)$ and
its solution $d_{\pm}$ is fixed, up to a constant factor, by the
boundary condition at the origin. The first two components of the
functions $F_{\pm}$ can be expressed in terms of $d_{\pm}$ so that
\begin{equation}
F_{\pm}\left(E,r\right)=\left(\begin{array}{c}
f_{\pm}\left(E,r\right)\\
h_{\pm}\left(E,r\right)\\
d_{\pm}\left(E,r\right)
\end{array}\right)=\left(\begin{array}{c}
\frac{2Md_{\pm}\left(E,r\right)}{r\left(V\left(r\right)-E\mp i0\right)}\\
\frac{2\partial_{r}d_{\pm}\left(E,r\right)}{V\left(r\right)-E\mp i0}\\
d_{\pm}\left(E,r\right)
\end{array}\right)\,,\label{eq:F+-}
\end{equation}
where we assume that the limit $\epsilon\to\pm0$ is already performed.
On the interval $\left(0,r_{\star}\left(E\right)\right)$ the functions
$F_{+}$ and $F_{-}$ coincide. Without loss of generality, we can
choose them to be real on this interval. Then, obviously, $F_{+}\left(E,r\right)=F_{-}^{*}\left(E,r\right)$
on the whole interval $\left(0,\infty\right)$. On the interval $\left(r_{\star}\left(E\right),\infty\right)$
the functions $F_{\pm}$ gain imaginary parts which is clearly seen
from the asymptotics of the functions $d_{\pm}$ in the vicinity of
$r_{\star}$:\begin{widetext}
\begin{eqnarray}
d_{\pm}\left(E,r\right) & \approx & b_{1}\left(1+\frac{M^{2}\left(1-r/r_{\star}\right)^{2}}{2}\ln\left(\frac{V\left(r\right)-E\mp i0}{\left|V^{\prime}\left(r_{\star}\right)\right|r_{\star}}\right)+\beta\left(1-r/r_{\star}\right)^{2}\right)\nonumber \\
 & \approx & b_{1}\left(1+\frac{M^{2}\left(1-r/r_{\star}\right)^{2}}{2}\ln\left(\frac{r_{\star}-r}{r_{\star}}\mp i0\right)+\beta\left(1-r/r_{\star}\right)^{2}\right)\,.
\end{eqnarray}
The prescription $\pm i0$ determines the choice of the logarithm
branch for $r>r_{\star}$, so that
\begin{equation}
d_{\pm}\left(E,r\right)\approx b_{1}\left(1+\frac{M^{2}\left(1-r/r_{\star}\right)^{2}}{2}\ln\left|\frac{r_{\star}-r}{r_{\star}}\right|+\left(\beta\mp i\frac{\pi M^{2}}{2}\theta\left(r-r_{\star}\right)\right)\left(1-r/r_{\star}\right)^{2}\right)\,.\label{eq:d_+-}
\end{equation}
\end{widetext}Using this formula, as well as the identity
\begin{equation}
\frac{1}{V\left(r\right)-E\mp i0}=\Pv\frac{1}{V\left(r\right)-E}\pm i\pi\delta\left(V\left(r\right)-E\right)\,,\label{eq:PV+delta}
\end{equation}
one can check that the real part of $F_{\pm}$ is proportional to
the function $F_{1}$, Eq. \eqref{eq:F1}, and the imaginary part
of $F_{\pm}$ is proportional to the function $F_{2}$, Eq. \eqref{eq:F2}.

\paragraph*{Orthonormality and the dual basis.}

For the potential decreasing faster than $1/r$, the asymptotics of
the functions $d_{\pm}$ at $r\to\infty$ has the form
\begin{equation}
d_{\pm}\left(E,r\right)\to\frac{c}{2\sqrt{r}}\left(e^{\mp i\left(Er+\varphi\right)/2}+\gamma e^{\pm i\left(Er+\varphi\right)/2}\right)\,,\label{eq:d asymptotics}
\end{equation}
where $c$, $\gamma$, and $\varphi$ are, in general, some real-valued
functions of the energy. For the potential decreasing at $r\to\infty$
as $\alpha/r$, one should perform the replacement $\varphi\to\varphi-\alpha\ln Er$.
We choose the overall normalization constant $c$ to be equal to unity,
so that the asymptotics of the functions $F_{\pm}$ has the form\begin{widetext}
\begin{equation}
F_{\pm}\left(E,r\right)\to\frac{1}{2\sqrt{r}}\left[\left(\begin{array}{c}
0\\
\pm i\\
1
\end{array}\right)e^{\mp i\left(Er+\varphi\right)/2}+\gamma\left(\begin{array}{c}
0\\
\mp i\\
1
\end{array}\right)e^{\pm i\left(Er+\varphi\right)/2}\right]\,.\label{eq:F+- asymptotics}
\end{equation}
\end{widetext} Then the functions $F_{\pm}$ satisfy the relation
\begin{gather}
\intop_{0}^{\infty}dr\, rF_{\sigma}^{\dagger}\left(E,r\right)F_{\sigma^{\prime}}\left(E^{\prime},r\right)=2\pi\delta\left(E-E^{\prime}\right)N_{\sigma\sigma^{\prime}}\,,\label{eq:FOrthoNorm}\\
N_{++}=N_{--}=\frac{1+\gamma^{2}}{2}+\frac{4\pi M^{2}\left|b_{1}\right|^{2}}{r^{\star}\left|V^{\prime}\left(r_{\star}\right)\right|}\,,\\
N_{+-}=N_{-+}=\gamma\,,
\end{gather}
where $b_{1}=d_{\pm}\left(E,r_{\star}\right)$, see Eq. \eqref{eq:d_+-}.
There is one subtle relation between constants $b_{1}$ and $\gamma$,
which follows from the conservation of the total radial current $J_{r}\left(r\right)=2\pi rj_{r}$,
where $j_{r}$ is defined in Eq. \eqref{eq:Psi0Current}. Namely,
using the equality $J_{r}\left(r\to r_{\star}+0\right)=J_{r}\left(r\to\infty\right)$,
we obtain
\begin{equation}
\frac{4\pi M^{2}\left|b_{1}\right|^{2}}{r^{\star}\left|V^{\prime}\left(r_{\star}\right)\right|}=\frac{1-\gamma^{2}}{2}\,,\label{eq:CurentConservationRel}
\end{equation}
so that
\begin{equation}
N_{++}=N_{--}=1\,.\label{eq:N++N--}
\end{equation}
Note that $J_{r}\neq0$ in the region $r>r_{\star}$. The existence
of the solutions with the nonzero radial current is the consequence
of the spectrum degeneracy. For $r<r_{\star}$we have $J_{r}=0$.
It may seem that such a behaviour contradicts the continuity equation
at $r=r_{\star}$. However, the density $\rho=F^{\dagger}F$ is ill-defined
at $r=r_{\star}$, so it does not make sense to consider the continuity
condition for $j_{r}$ at this point.

Since the wave functions $F_{\pm}\left(E,r\right)$ are not orthogonal
to each other, it is convenient to introduce the dual basis functions
$G_{\pm}\left(E,r\right)$ being the linear combinations of $F_{\pm}\left(E,r\right)$
and satisfying the relations:
\begin{equation}
\intop_{0}^{\infty}dr\, rG_{\sigma}^{\dagger}\left(E,r\right)F_{\sigma^{\prime}}\left(E^{\prime},r\right)=2\pi\delta\left(E-E^{\prime}\right)\delta_{\sigma\sigma^{\prime}}\,.
\end{equation}
From Eqs. \eqref{eq:FOrthoNorm}, \eqref{eq:N++N--} we have
\begin{equation}
G_{\pm}\left(E,r\right)=\frac{F_{\pm}-\gamma F_{\mp}}{1-\gamma^{2}}\,.
\end{equation}
Using Eq. \eqref{eq:F+- asymptotics}, we find that the large-distance
asymptotics of $G_{\pm}$ corresponds to the convergent/divergent
spherical wave, respectively:
\begin{equation}
G_{\pm}\left(E,r\right)\to\frac{e^{\mp i\left(Er+\varphi\right)/2}}{2\sqrt{r}}\left(\begin{array}{c}
0\\
\pm i\\
1
\end{array}\right)\,.\label{eq:G+- asymptotics}
\end{equation}
This remarkable property is important in the consideration of the
time-dependent problem.

\section{Time-dependent problem}

The stationary solutions for $M\neq0$, derived in the previous section,
look very unusual due to a singular behaviour at $r=r_{\star}$. This
behaviour should be reflected in the time evolution of wave packets.
Note that Eq. \eqref{eq:F equation} has the form resolved with respect
to the derivative $\partial_{t}F$. Besides, the coefficients of the
differential operator $\hat{H}_{r}$ in the right-hand side are smooth
functions on the interval $\left(0,\infty\right)$. So, for suitable
initial and boundary conditions, the problem of finding the solution
$F\left(t,r\right)$ is well-posed.

We choose the initial conditions as

\begin{eqnarray}
F\left(0,r\right) & = & \frac{C}{\sqrt{r}}\left(\begin{array}{c}
0\\
i\\
1
\end{array}\right)e^{-i\frac{E_{0}r}{2}}\Omega\left(\frac{r-r_{0}}{\Delta}\right)\,,\label{eq:WavePacket}\\
\Omega\left(x\right) & = & \left(1-x^{2}\right)^{2}\theta\left(1-x^{2}\right),
\end{eqnarray}
where $C$ is the normalization constant, determined by the relation
$\int_{0}^{\infty}dr\, r\left|F\right|^{2}=1$. This form corresponds
to the scattering problem setup and describes the wave packet with
the average energy $E_{0}$ and the width $\Delta$, moving from the
large distance $r_{0}$ towards the origin (cf. Eq. \eqref{eq:F+- asymptotics}).
We assume that $r_{0}\gg\Delta\gg\left|E_{0}\right|^{-1}$, i.e.,
the packet width in the coordinate space is small compared to the
average value of $r$ , and the width in the momentum space is small
compared to the average value of $E$. If the potential $V\left(r\right)$
is a localized function falling off at $r\sim R$, then we also assume
that $\Delta\gg R$. Performing the numerical integration of Eq. \eqref{eq:F equation}
over $t$, we find $F\left(t,r\right)$. The initial packet with the
energy $E_{0}$ well above $V_{\mathrm{max}}$ (when $E_{0}-V_{\mathrm{max}}\gg1/\Delta$)
or well below zero ($-E_{0}\gg1/\Delta$) moves with the speed $2v_{F}$
($2$ in our units), comes to small distances, reflects, and goes
away. The norm of the outgoing packet is the same as that of the incoming
one. This behaviour looks very similar to that of the wave packet
obeying the massless Dirac equation in the central external field.
The evolution of the wave packet with the energy $E_{0}$ deep inside
the interval $\left(0,V_{\mathrm{max}}\right)$ is essentially different.
During the scattering process a narrow peak develops at $r_{\star}\left(E_{0}\right)$.
The form of the peak stabilizes at large time. The norm of the outgoing
packet (corresponding to the reflected particles) is less than that
of the incoming one. However, the total norm is conserved due to the
additional contribution of the peak at finite distances (corresponding
to the ``adhered'' particles).
\begin{figure*}
\hspace{-30pt}\includegraphics[width=8cm]{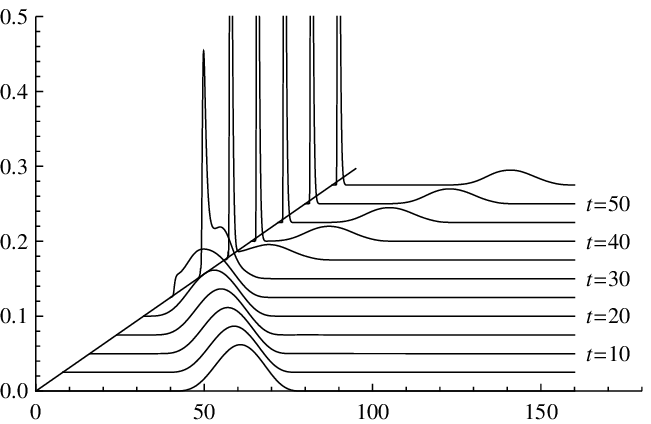}\begin{picture}(0,0)
\put(-225,155){$r |F|^2$}
\put(5,10){$r$}
\put(-100,90){$t$}
\end{picture}\hspace{-83pt}\raisebox{101pt}{\includegraphics{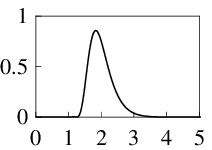}}\hspace{50pt}\includegraphics[width=8cm]{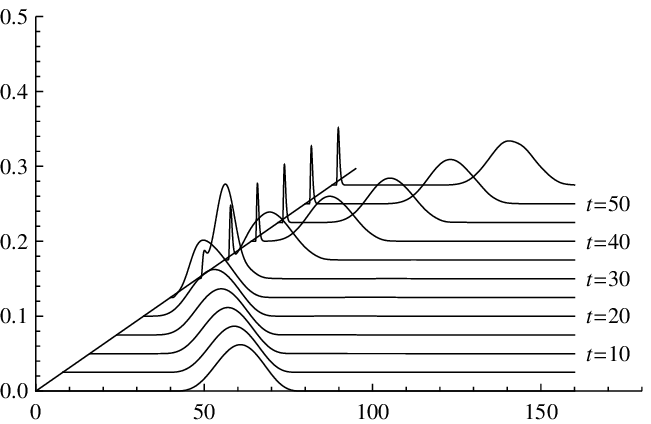}\begin{picture}(0,0)
\put(-225,155){$r |F|^2$}
\put(5,10){$r$}
\put(-100,90){$t$}
\end{picture}\hspace{-83pt}\raisebox{101pt}{\includegraphics{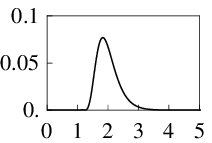}}\caption{Time evolution of the density, corresponding to the wave packet \eqref{eq:WavePacket},
in the Coulomb potential $V\left(r\right)=\alpha/r$. The parameters
are , $E_{0}=1$, $r_{0}=60$, $\Delta=20$ , $\alpha=2$, $M=1$
(left) and $M=2$ (right). Insets: the form of the peak at large time.
\label{fig:TimeEvolution}}
\end{figure*}
To demonstrate this behaviour, the time evolution of the wave packet
\eqref{eq:WavePacket} in the Coulomb potential $V\left(r\right)=\alpha/r$
is shown in Fig. \ref{fig:TimeEvolution}. One can see that, after
reflection, a narrow peak appears at $r_{\star}=\alpha/E_{0}$.

\paragraph*{Decomposition method.}

In order to gain deeper insight into this behaviour, let us derive
the time evolution of the wave packet using the decomposition of the
initial wave packet over the stationary wave functions. The decomposition
has the form\begin{widetext}

\begin{eqnarray}
F\left(t,r\right) & = & \intop_{0}^{V_{\mathrm{max}}}\frac{dE}{2\pi}e^{-iEt}\left[C_{+}\left(E\right)F_{+}\left(E,r\right)+C_{-}\left(E\right)F_{-}\left(E,r\right)\right]\nonumber \\
 &  & +\intop_{-\infty}^{0}\frac{dE}{2\pi}C\left(E\right)e^{-iEt}F_{1}\left(E,r\right)+\intop_{V_{\mathrm{max}}}^{\infty}\frac{dE}{2\pi}C\left(E\right)e^{-iEt}F_{1}\left(E,r\right)\,.\label{eq:decomposition}
\end{eqnarray}
\end{widetext}The coefficients $C\left(E\right)$ and $C_{\pm}\left(E\right)$
have the form
\begin{eqnarray}
C\left(E\right) & = & \intop_{0}^{\infty}dr\, rF_{1}^{\dagger}\left(E,r\right)F\left(0,r\right),\label{eq:C(E)}\\
C_{\pm}\left(E\right) & = & \intop_{0}^{\infty}dr\, rG_{\pm}^{\dagger}\left(E,r\right)F\left(0,r\right).\label{eq:C+-(E)}
\end{eqnarray}
Note that the coefficients $C_{\pm}\left(E\right)$ in front of $F_{\pm}\left(E,r\right)$
in the decomposition \eqref{eq:decomposition} are expressed via the
overlap integrals of $F\left(0,r\right)$ with the elements $G_{\pm}^{\dagger}\left(E,r\right)$
of the dual basis. Let us consider the decomposition of the wave packet
\eqref{eq:WavePacket} with the energy $E_{0}$ deep inside the interval
$\left(0,V_{\mathrm{max}}\right)$, when $V_{\mathrm{max}}-E_{0}\gg1/\Delta$
and $E_{0}\gg1/\Delta$. In this case the main contribution to the
integrals in Eqs. \eqref{eq:C(E)} and \eqref{eq:C+-(E)} comes from
large distances $r\sim r_{0}$. Therefore, for the calculation of
the coefficients $C\left(E\right)$ and $C_{\pm}\left(E\right)$,
we can use the large-$r$ asymptotics \eqref{eq:F1 asymptotics} and
\eqref{eq:F+- asymptotics}. We obtain that $C\left(E\right)$ and
$C_{-}\left(E\right)$ are suppressed due to the fast oscillations
of the integrands, and we can omit the corresponding contributions
in Eq. \eqref{eq:decomposition}. The coefficient $C_{+}\left(E\right)$
has the form\begin{widetext}
\begin{align}
C_{+}\left(E\right) & =\tilde{C}\left(E\right)\exp\left(iEr_{0}/2\right)\tilde{\Omega}\left(\frac{\left(E-E_{0}\right)\Delta}{2}\right)\,,\\
\tilde{\Omega}\left(q\right) & =\int dx\exp\left(iqx\right)\Omega\left(x\right)=\frac{16}{q^{5}}\left[\left(3-q^{2}\right)\sin q-3q\cos q\right]\,,\label{eq:tildeOmega}\\
\tilde{C}\left(E\right) & =\frac{\sqrt{\Delta/2}}{\sqrt{\int dx\Omega^{2}\left(x\right)}}e^{-i\left(E_{0}r_{0}-\varphi\left(E\right)\right)/2}=\sqrt{\frac{315\Delta}{512}}e^{-i\left(E_{0}r_{0}-\varphi\left(E\right)\right)/2}\,.\label{eq:tildeC}
\end{align}
\end{widetext}The function $\tilde{\Omega}\left(\frac{\left(E-E_{0}\right)\Delta}{2}\right)$
is peaked around $E=E_{0}$ with the characteristic width $1/\Delta$,
while $\tilde{C}\left(E\right)$ is some slowly varying function of
the energy. Therefore, we can represent the function $F\left(t,r\right)$
as
\begin{equation}
F\left(t,r\right)=\tilde{C}\left(E_{0}\right)\!\!\!\!\intop_{0}^{V_{\mathrm{max}}}\!\!\frac{dE}{2\pi}e^{-iE\tau}\tilde{\Omega}\left(\frac{\left(E-E_{0}\right)\Delta}{2}\right)F_{+}\left(E,r\right),
\end{equation}
where $\tau=t-r_{0}/2$.

Let us demonstrate now that this decomposition leads to the appearance
of the peak in the vicinity of $r=r_{\star}\left(E_{0}\right)$ at
large $t$. For this purpose we consider the asymptotic form of $F\left(t,r\right)$
for $t$ satisfying the condition $\left|2t-r_{0}\right|\gg\Delta$
and for $r$ obeying the condition $r_{\star}\left|V^{\prime}\left(r_{\star}\right)\right|\left|r-r_{\star}\right|\ll1$.
Keeping in $F_{+}\left(E,r\right)$, Eq. \eqref{eq:F+-} , only the
singular component
\begin{eqnarray}
f_{+}\left(E,r\right) & = & \frac{2Md_{+}\left(E,r\right)}{r\left(V\left(r\right)-E-i0\right)}\,,
\end{eqnarray}
and using the identity \eqref{eq:PV+delta}, we obtain\begin{widetext}

\begin{align}
f\left(t,r\right) & \approx\frac{2M\tilde{C}\left(E_{0}\right)}{r}\intop_{0}^{V_{\mathrm{max}}}\frac{dE}{2\pi}e^{-iE\tau}\tilde{\Omega}\left(\frac{\left(E-E_{0}\right)\Delta}{2}\right)\left(\Pv\frac{1}{V-E}+i\pi\delta\left(V-E\right)\right)d_{+}\left(E,r\right),
\end{align}
where $V=V\left(r\right)$. Passing to the variable $\varepsilon=E-V$
we have
\begin{equation}
f\left(t,r\right)\approx\frac{2M\tilde{C}\left(E_{0}\right)}{r}e^{-iV\tau}\intop_{L_{1}}^{L_{2}}\frac{d\varepsilon}{2\pi}e^{-i\tau\varepsilon}\tilde{\Omega}\left(\frac{\Delta\left(V-E_{0}+\varepsilon\right)}{2}\right)\left(i\pi\delta\left(\varepsilon\right)-\Pv\frac{1}{\varepsilon}\right)d_{+}\left(V+\varepsilon,r\right),\label{eq:f(t,r)}
\end{equation}
where $L_{1}=-V$, $L_{2}=V_{\mathrm{max}}-V$. Since $\left|\tau\right|\gg\Delta$
and $\left|\tau L_{1,2}\right|\gg1$, we can write Eq. \eqref{eq:f(t,r)}
as
\begin{equation}
f\left(t,r\right)\approx\frac{2M\tilde{C}\left(E_{0}\right)}{r}e^{-iV\tau}\tilde{\Omega}\left(\frac{\Delta\left(V-E_{0}\right)}{2}\right)d_{+}\left(V,r\right)\intop_{-\infty}^{\infty}\frac{d\varepsilon}{2\pi}e^{-i\tau\varepsilon}\left(i\pi\delta\left(\varepsilon\right)-\Pv\frac{1}{\varepsilon}\right)\,.
\end{equation}
The remaining integral is equal to $i\theta\left(\tau\right)$, and
finally we come to the asymptotics of $f\left(t,r\right)$ at $\ensuremath{\left|\tau\right|=\left|t-r_{0}/2\right|\gg\Delta}$:
\begin{equation}
f\left(t,r\right)\approx\frac{2iM\tilde{C}\left(E_{0}\right)}{r}e^{-iV\tau}\tilde{\Omega}\left(\frac{\Delta\left(V-E_{0}\right)}{2}\right)d_{+}\left(V,r\right)\theta\left(\tau\right)\,.\label{eq:f(t,r) asymptotic}
\end{equation}
\end{widetext}Note that the right-hand side of Eq. \eqref{eq:f(t,r) asymptotic}
vanishes for $\tau\ll-\Delta$ , and the leading asymptotics of $F\left(t,r\right)$
comes from the contribution of the nonsingular terms. However, we
can claim that this asymptotics is not peaked in the vicinity of $r_{\star}\left(E_{0}\right)$.
For $\tau\gg\Delta$, the density $\left|F\right|^{2}\approx\left|f\right|^{2}$
is independent of $\tau$ and peaked, due to the factor $\tilde{\Omega}$
in Eq. \eqref{eq:f(t,r) asymptotic}, with the characteristic width
$\delta\sim1/\left|V^{\prime}\left(r_{\star}\right)\Delta\right|$.
Thus, we have demonstrated that the decomposition method leads to
the appearance of the peak at large time, which is in agreement with
the result of direct numerical solution of the differential equation.
For the Coulomb potential, we have also checked numerically that the
time evolution, obtained by the decomposition method, coincides with
that obtained by the direct numerical solution of the differential
equation, see Fig. \ref{fig:TimeEvolution}.

\paragraph*{Adhesion coefficient.}

Let us consider the quantity
\begin{equation}
A=\lim_{t\to+\infty}\intop_{0}^{L}dr\, r\left|F\left(t,r\right)\right|^{2}\,.
\end{equation}
 The upper limit $L$ in this formula is some fixed parameter obeying
the condition $L\gg r_{\star}\left(E_{0}\right)$. The quantity $A$
is the adhesion coefficient, i.e., the probability for the two particles
to remain at finite distances at large $t$. Using Eqs. \eqref{eq:f(t,r) asymptotic},
\eqref{eq:tildeOmega}, \eqref{eq:tildeC}, \eqref{eq:d_+-}, and
\eqref{eq:CurentConservationRel} we obtain
\begin{equation}
A\approx\lim_{t\to+\infty}\intop_{0}^{L}dr\, r\left|f\left(t,r\right)\right|^{2}\approx1-\gamma^{2}\,,\label{eq:AdhesionCoefficient}
\end{equation}
where $\gamma=\gamma\left(E_{0}\right)$. For the Coulomb potential
$V\left(r\right)=\alpha/r$, the dimensional arguments lead to the
independence of the quantity $\gamma$ of $\left|E_{0}\right|$.
\begin{figure}
\includegraphics[width=8cm]{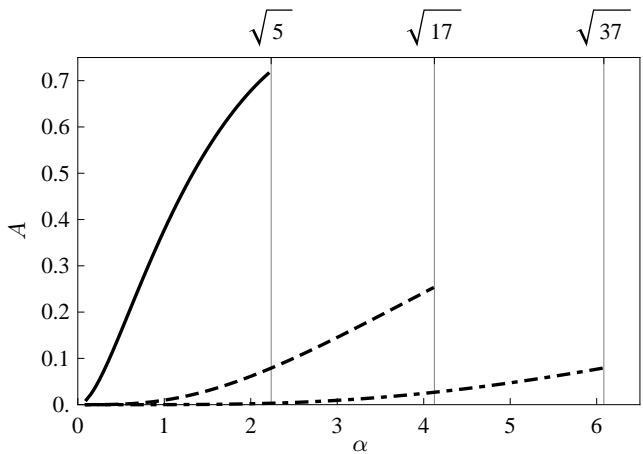}\begin{picture}(0,0)
\put(-240,75){\rotatebox{90}{$A$}}
\put(-110,-5){$\alpha$}
\end{picture}\caption{The adhesion coefficient $A$, Eq. \eqref{eq:AdhesionCoefficient},
in the Coulomb potential as a function of $\alpha$ for $M=1$ (solid
curve), $M=2$ (dashed curve), $M=3$ (dash-dotted curve). The vertical
lines correspond to the critical values $\alpha_{M}=\sqrt{1+4M^{2}}$.\label{fig:AdhesionCoefficient}}

\end{figure}
In Fig. \ref{fig:AdhesionCoefficient} the adhesion coefficient $A$
for the case of Coulomb potential is shown as a function of $\alpha$.
One can see that $A$ grows when $\alpha$ changes from $0$ to its
critical value $\alpha_{M}=\sqrt{1+4M^{2}}$ (when the parameter $\nu$
in Eq. \eqref{eq:d r=00003D0 asymptotics Coulomb} vanishes).

\paragraph*{Asymptotically localized state.}

The numerical solution of the differential equation \eqref{eq:F equation}
shows that at large $t$ and fixed $r$ the functions $h\left(t,r\right)$
and $d\left(t,r\right)$ vanish. Then, it follows from Eq. \eqref{eq:f equation}
that $f\left(t,r\right)\overset{t\to\infty}{\longrightarrow}e^{-iV\left(r\right)t}f_{0}\left(r\right)$,
where $f_{0}\left(r\right)$ is some function of $r$. This form of
the asymptotics is also in agreement with Eq. \eqref{eq:f(t,r) asymptotic}.
The function $f_{0}\left(r\right)$ is peaked at $r=r_{\star}$ and
depends on the form of the initial packet \eqref{eq:WavePacket}.
This asymptotics could be considered as a hint for the existence of
the normalizable solutions of Eq. \eqref{eq:F equation} with the
constant density $\rho\left(r\right)$=$\left|f_{0}\left(r\right)\right|^{2}$.
However, the equation \eqref{eq:F equation} has no solutions of the
form
\[
F\left(t,r\right)=e^{-iV\left(r\right)t}\begin{pmatrix}f_{0}\left(r\right)\\
0\\
0
\end{pmatrix}\,.
\]
Indeed, this form satisfies Eqs. \eqref{eq:f equation} and \eqref{eq:h equation},
but not \eqref{eq:d equation}. Instead, we search the asymptotics
of the solution of Eqs. \eqref{eq:f equation}-\eqref{eq:d equation}
as
\begin{equation}
F\left(t,r\right)=e^{-iV\left(r\right)t}\sum_{n=0}^{\infty}\begin{pmatrix}f_{n}\left(r\right)\\
h_{n}\left(r\right)\\
d_{n}\left(r\right)
\end{pmatrix}t^{-n}\,.\label{eq:F asymptotics}
\end{equation}
Substituting this form in Eq. \eqref{eq:F equation}, we obtain the
recurrence relations for $f_{n}\left(r\right)$, $h_{n}\left(r\right)$,
and $d_{n}\left(r\right)$ which can be used to express the asymptotics
\eqref{eq:F asymptotics} via one function $f_{0}\left(r\right)$.
In the leading order we obtain

\begin{align}
f\left(t,r\right) & =e^{-iV\left(r\right)t}\left[f_{0}\left(r\right)+O\left(t^{-2}\right)\right]\,,\label{eq:fhd asymptotics}\\
h\left(t,r\right) & =\frac{e^{-iV\left(r\right)t}}{t}\left[\frac{iMf_{0}\left(r\right)}{rV^{\prime}\left(r\right)}+O\left(t^{-1}\right)\right]\,,\\
d\left(t,r\right) & =\frac{e^{-iV\left(r\right)t}}{t^{3}}\left[-\frac{iMf_{0}\left(r\right)}{2rV^{\prime2}\left(r\right)}+O\left(t^{-1}\right)\right]\,.\label{eq:d(t,r) asymptotics}
\end{align}
Though the function $h\left(t,r\right)$ vanishes at $t\to\infty$
, its derivative $\partial_{r}h\left(t,r\right)$ does not vanish,
$\left|\partial_{r}h\left(t,r\right)\right|\to\left|Mf\left(r\right)/r\right|$.
Due to this behaviour of $h\left(t,r\right)$, the equation \eqref{eq:d equation}
is now satisfied. We see that the asymptotics \eqref{eq:fhd asymptotics}-\eqref{eq:d(t,r) asymptotics}
is in agreement with the behaviour observed in the numerical solution
of the differential equation. Note that the asymptotics \eqref{eq:fhd asymptotics}-\eqref{eq:d(t,r) asymptotics}
leads to the vanishing radial and azimuthal current, Eq. \eqref{eq:Psi0Current}.

\section{Conclusion}

In the present paper, we have considered a model of the electron-electron
interaction in graphene, based on the hamiltonian \eqref{eq:H}. Despite
the simplicity of the model, it leads to a very unusual dynamics.
We have shown, both numerically and analytically, that in the process
of the wave packet scattering the asymptotically localized state appears,
see Fig. \eqref{fig:TimeEvolution} and Eq. \eqref{eq:f(t,r) asymptotic}.
From the point of view of the outside observer, the scattering seems
to be inelastic. The origin of this phenomenon is traced back to two-fold
degeneracy of the spectrum of hamiltonian $H_{r}$, Eq. \eqref{eq:Hr},
at $0<E<V_{\mathrm{max}}$, Eqs. \eqref{eq:F1} and \eqref{eq:F2}.
This degeneracy is related to the degeneracy of the derivative matrix
in $\hat{H}_{r}$. The results obtained are valid for any smooth monotonically
decreasing vanishing potential. Though we did not take into account
the Fermi statistics of the interacting electrons, the requirement
of the Fermi statistics can be satisfied by appropriate choice of
the spin part of the wave function.

For simplicity, we considered the states with definite value of $J^{z}$.
It is obvious, that the observed phenomenon (the appearance of the
asymptotically localized state) also retains for any superposition
of the states with different values of $J^{z}$. Our consideration
is not directly applicable to the case $\mathbf{P}\neq0$, but the
appearance of the asymptotically localized state is likely to take
place also in this case because the derivative matrix in $H_{V}$,
Eq. \eqref{eq:H}, is also degenerate.

We emphasize that this simple model does not take into account the
existense of the electrons below Fermi surface. The effect of such
electrons is the appearance of virtual electron-hole pairs in the
intermediate states. While the exact account of this effect is hardly
possible, its qualitative consideration is very important and will
be presented elsewhere. It seems that the effect of the electrons
below Fermi surface is small at least in the case when $\delta E_{F}<0$
and $\left|\delta E_{F}\right|\gg E_{0}$, where $\delta E_{F}$ is
the difference between the Fermi energy and the energy of the Dirac
point. Therefore, it should be possible to observe the appearance
of the asymptotically localized state in the experiment. Note that
the differential equation \eqref{eq:H[P=00003DP0]} does not take
into account the effect of the finite lattice scale $l_{\mathrm{0}}\sim0.14\mbox{nm}$,
and valid if the wave function changes slowly on this scale. It gives
us two conditions: the width of the localized state should be much
larger than the lattice scale, and the variation of the phase on the
lattice scale should be small compared to unity. The first condition
gives the constraint on the energy dispersion $\delta E$ in the wave
packet:
\[
\delta E\gg\left|V^{\prime}\left(r_{\star}\right)\right|l_{0}\,.
\]
It follows from Eq. \eqref{eq:f(t,r) asymptotic} that, at large time,
the phase of the wave function varies significantly on the lattice
scale. The second condition gives us the lower bound $\tau_{0}$ of
the life time of the localized state:
\[
\tau_{0}\sim\frac{1}{l_{0}\left|V^{\prime}\left(r_{\star}\right)\right|}\gg1/\delta E\,,
\]
which means that the localized state lives long enough. For the Coulomb
potential, these two conditions read
\begin{eqnarray*}
\delta E & \gg & \frac{E_{0}^{2}l_{0}}{\alpha},\\
\tau_{0} & \sim & \frac{\alpha}{l_{0}E_{0}^{2}}\,.
\end{eqnarray*}
Note that the first condition is compatible with the condition $\delta E\ll E_{0}$
provided that $E_{0}$ is sufficiently small. For instance, for $E_{0}\sim1\mathrm{meV}$
and $\alpha\sim1$, we have quite large time $\tau_{0}\sim1\mathrm{\mu s}$.
Therefore, if the model considered is relevant to the electron-electron
interaction in graphene, one may hope to observe the long-lived localized
states in the experiment.

\paragraph*{Acknowledgements}

This work was supported by Federal special-purpose program \textquotedblleft{}Scientific
and scientific-pedagogical personnel of innovative Russia\textquotedblright{},
RFBR grant No. 11-02-00220. The work of I.S.T. was also supported
by the \textquotedblleft{}Dynasty\textquotedblright{} foundation.

\appendix

\bibliographystyle{apsrev}
%\bibliography{Refs}

\end{document}